\documentclass
[superscriptaddress,twocolumn,showpacs,amsmath,amssymb]{revtex4}
\usepackage[dvips]{graphicx}
\usepackage{amsmath,amsthm,amssymb,bm}
\newcommand{\bi}[1]{\ensuremath{\boldsymbol{#1}}}
\newcommand{\unit}[1]{\ensuremath{\mathrm{#1}}}

\def \beq{\begin{equation}}
\def \eeq{\end{equation}}
\def \beqa{\begin{eqnarray}}
\def \eeqa{\end{eqnarray}}

\def \Hami{Hamiltonian}
\def \twop{2{\it p} }
\def \etal{{\it et al.}}

\begin{document}

\title{Time-dependent approach to many-particle tunneling in one-dimension}

\author{Takahito Maruyama}
\altaffiliation[Present address:]{
Fusion Research and Development Directorate, Japan
Atomic Energy Agency (JAEA), Naka, Ibaraki 311-0193, Japan}
\affiliation{
Department of Physics,  Tohoku University,  Sendai 980-8578,  Japan}

\author{Tomohiro Oishi}
\author{Kouichi Hagino}
\affiliation{
Department of Physics,  Tohoku University,  Sendai 980-8578,  Japan}
\author{Hiroyuki Sagawa}
\affiliation{
Center for Mathematics and Physics, University of Aizu,
Aizu-Wakamatsu, Fukushima 965-8560, Japan}
\affiliation{
RIKEN Nishina Center, Wako 351-0198, Japan}


\begin{abstract}
Employing the time-dependent approach, we investigate a quantum 
tunneling decay of many-particle systems. 
We apply it to 
a one-dimensional three-body problem with a heavy core 
nucleus and two valence protons. 
We calculate the decay width for two-proton emission 
from the survival probability, which 
well obeys the exponential decay-law after a sufficient time. 
The effect of the correlation between the two emitted protons 
is  also studied by observing 
the time evolution 
of the two-particle density 
distribution. 
It is shown that the pairing correlation significantly 
enhances the probability for 
the simultaneous diproton decay. 
\end{abstract}


\maketitle

\section{INTRODUCTION}

The quantum tunneling of a system with intrinsic degrees of 
freedom, or of many particles, is an important subject of modern 
physics\cite{B61,CL83,Flam05,Flam07,Ahs10,ShoSho11}. 
In nuclear physics, 
typical examples include heavy-ion fusion reactions at 
subbarrier energies\cite{BT98,DHRS98}, 
spontaneous fission\cite{VH74}, alpha and heavy-cluster 
decays\cite{D10}, and 
stellar nucleosynthesis\cite{BD04,TN09}. 
In heavy-ion fusion reactions, for instance, it has been well recognized that 
the couplings of the 
relative motion between the colliding nuclei to several 
collective motions enhance the tunneling probability of the 
Coulomb barrier, thus increasing the fusion cross sections, by several orders 
of magnitude as compared to a prediction of a simple 
potential model \cite{BT98,DHRS98}. 
Nevertheless, 
it has still been a challenging problem to understand the many-particle 
tunneling from a fully microscopic view. 
For instance, even though 
there have been 
several attempts 
\cite{D10,SKT70, TonAri79, VarLio94}, 
alpha decays have not fully been understood microscopically with 
sufficient accuracy. 

Recently, 
two-proton (2{\it p}) radioactivities have been experimentally 
observed for 
a few proton-rich nuclei 
outside the proton drip-line, 
such as 
$^{45}$Fe \cite{Pfu02,Gio02,Mie07} 
and $^6$Be \cite{OVB89,Gri01,Gri09}, 
and have attracted much attention
\cite{PKGR12, BP08,G09,GWM09} 
in connection to {\it e.g.,} 
the dinucleon 
correlations \cite{MMS05,Mat06,HS05,OHS10}. 
This is a phenomenon of 
spontaneous emission of two valence protons from 
the parent proton-rich nuclei, in which the emission of one 
proton is energetically forbidden. 
Notice that an analogous process of the two-proton radioactivity, 
that is, a two-neutron decay has also been observed recently 
for $^{16}$Be\cite{Spy12}. 
These two-nucleon emission decays may provide a useful testing ground of 
many-particle tunneling theories. 

A primary task for a many-particle tunneling decay is to investigate 
the effect of interaction or correlation among emitted particles 
on decay properties such as the decay width 
and the survival probability. 
In the case of \twop decay, this corresponds to the pairing 
correlation between the valence protons. 
Because of the 
strong pairing correlation in proton-rich nuclei, 
\twop emitters are expected to have an even number of protons outside the 
proton drip-line. 
Incidentally, it has been well known that 
the pairing correlation plays an important role 
in two-neutron transfer reactions \cite{GLV12,SLBW12,PBM11}. 

The quantum tunneling decay phenomena can be studied either with 
the time-independent approach \cite{D10,G09,GWM09}
or the time-dependent approach 
\cite{Car94,Car98,Car99A,Car99B,Car00,GCML11}. 
In the time-independent approach, one seeks {\it e.g.,} 
a Gamow state, which 
is a purely outgoing wave outside the barrier. 
The imaginary part of the energy of the Gamow state is related to the 
decay width, while the real part corresponds to the resonance energy. 
On the other hand, in the time-dependent approach, 
one first modifies the potential barrier so that 
the initial state can be prepared as a bound state of 
a confining potential. 
The confining potential is then suddenly changed to the original barrier, and 
the initial state evolves in time. The decay width can be obtained from 
the survival probability of the initial state. 
An advantage of the time-independent approach is that the decay width 
can be calculated with high accuracy even when the decay width is extremely 
small \cite{DE00}. An advantage of the time-dependent approach, on the 
other hand, is that it provides an intuitive way to understand the 
tunneling decay, even though it may be difficult to apply it to a 
situation with 
an extremely small decay width. 
This approach may provide a useful means to explore the mechanism of 
many-particle 
tunneling decay, though 
it has so far been applied only to two-body 
decay phenomena, such as $\alpha$ decays and one-proton decays 
\cite{Car94,Car98,Car99A,Car99B,Car00}. 

In this paper, we extend the time-dependent approach to two-proton 
emissions and discuss the dynamics of a two-particle tunneling decay. 
To this end, we employ the one-dimensional three-body 
model \cite{HVPS11}, which consists of a heavy core nucleus and two 
valence protons. 
We solve the time-dependent Schr\"odinger 
equation by expanding the wave function on 
a basis with 
time-dependent expansion coefficients. The decay width is then 
defined from the survival probability. 
We shall study the time evolution of the survival probability 
as well as the density distribution. We shall also discuss the role of 
pairing correlation in the two-proton decay. 

The paper is organized as follows. 
In Sec. II we detail our formalism for the time-dependent 
approach to the two-proton decays. 
We show the results of the calculations in Sec. III. 
It will be shown that the decay width converges to a constant 
value after a sufficient 
time evolution, indicating that the decay rate follows the exponential law. 
With this method, the time-evolution of a quasi-stationary 
\twop state, namely the density and the flux distributions, 
can be visualized. 
Using them, 
we shall discuss an important role of the 
pairing interaction between the two protons in the decay process. 
Finally we summarize the paper in Sec. IV. 

\section{FORMALISM}

\subsection{One-dimensional three-body model}

We consider a one-dimensional three-body system with 
two valence protons and the core-nucleus 
whose atomic and mass numbers are $Z_c$ and $A_c$, respectively. 
Neglecting the recoil kinetic energy of the core nucleus, 
the three-body Hamiltonian reads \cite{HVPS11},
\begin{eqnarray}
H &=& h(x_1) + h(x_2) + v_{\rm pp} (x_1,x_2),
\label{eq:Hami} \\
h(x) &=& -\frac{{\hbar}^2}{2m } \frac{d^2}{dx^2} + V(x),
\label{eq:sph}
\end{eqnarray}
where $m$ is the nucleon mass, 
$x_1$ and $x_2$ are the coordinates of the valence protons 
with respect to the core nucleus. 
$V(x)$ is the potential between a valence proton and the core, 
whereas $v_{\rm pp} (x_1,x_2)$ is the interaction between the 
two valence protons. 
The core-proton potential, 
\begin{eqnarray}
V(x)=V_{\rm nucl}(x)+V_{\rm coul}(x), 
\label{eq:core-p}
\end{eqnarray}
consists of the nuclear part 
$V_{\rm nucl}$ 
and the Coulomb part 
$V_{\rm coul}$. 
For the nuclear part, we take a Woods-Saxon form, 
\begin{equation}
V_{\rm nucl}(x) = -\frac{V_0}{1+e^{(|x|-R)/a}}. 
\label{eq:Vcp_N} 
\end{equation}
For the Coulomb part, 
we employ a soft-core Coulomb potential\cite{PGB91,GE92,LGE00}, that is, 
\begin{equation}
V_{\rm coul}(x) = \frac{Z_ce^2}{\sqrt{b^2+x^2}}.
\label{eq:Vcp_C}
\end{equation}
In this paper we take $V_0 = 46.5$ MeV. 
The radius $R$ and the surface diffuseness parameter $a$ 
in the Woods-Saxon potential, Eq. (\ref{eq:Vcp_N}), 
are taken to be $R=1.27A_c^{1/3}$ fm and $a=0.67 \; \unit{fm}$, 
respectively. 
We arbitrary take $A_c=60$ and $Z_c=30$ for 
the mass- and atomic- numbers of the core nucleus, 
while we use $b=2.0 \; \unit{fm}$ in the Coulomb interaction. 

The interaction $v_{\rm pp} (x_1,x_2)$ 
induces the pairing correlation between 
the two valence protons. 
In this work, we adopt a density-dependent 
contact interaction of the surface type\cite{HVPS11}, that is, 
\begin{equation}
v_{\rm pp}(x_1,x_2) = -g \left( 1-\frac{1}{1+e^{(|\bar{x}|-R)/a}} \right)
\delta(x_1-x_2),
\label{eq:vpair}
\end{equation}
where $g$ is the strength of the interaction and 
$\bar{x}=(x_1+x_2)/2$. 
The density 
dependence is introduced with the Woods-Saxon form 
($R$ and $a$ are the same as those in Eq.(\ref{eq:Vcp_N})). 
Notice that in the limit of $|\bar{x}| \rightarrow \infty $, this 
interaction becomes a pure contact interaction, $-g\delta(x_1-x_2)$. 
For simplicity, we neglect the Coulomb interaction between 
the two protons. We have confirmed that, as long as the one-dimensional 
three-body 
system is concerned,  
its effect on the decay properties can be well taken into account 
by somewhat reducing the strength $g$ 
(see also Refs. \cite{OHS10,OHS11,NY11}). 

It is important to notice that a one-dimensional delta function potential
$v(x)=-g\,\delta(x)$ always
holds a bound state at $E_{\rm pp}=-mg^2/4\hbar^2$ for a two-proton 
system
even with an infinitesimally
small attraction $g$ \cite{QM-book}.
This is in contrast to a three-dimensional system, in which a bound 
state exists only with a strong strength $g$ 
of an attractive contact interaction. 

\begin{figure}[t]
\begin{center}
\includegraphics[width=\hsize]{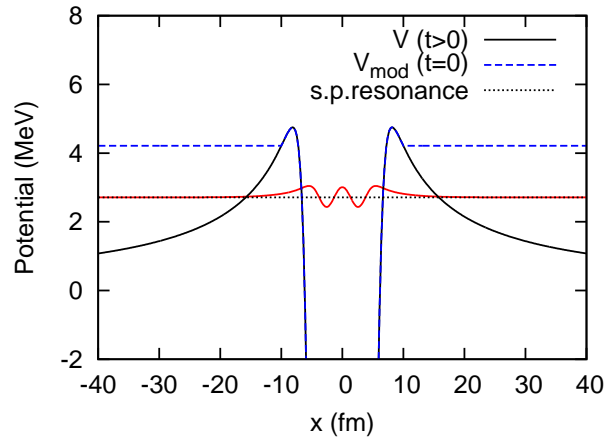}
\end{center}
\caption{
(Color online) The core-proton potential used 
in our calculations. 
The initial state at $t=0$ 
is constructed with a modified potential 
$V_{\rm mod}(x)$ given by Eq. \eqref{eq:vmod}, which is 
shown by the dashed line. 
For $t>0$, the potential is changed to the original 
potential, $V(x)$, given by Eq. \eqref{eq:core-p}, 
as shown by the solid line. 
We also show the wave function for the 
bound state of the modified potential 
at $\epsilon = 2.71$ MeV. }
\label{fig:fig1}
\end{figure}

\subsection{Time-dependent method}

In order to describe the \twop tunneling, we employ the time-dependent 
method. 
The first step is to prepare the initial \twop state which is 
confined inside the potential barrier, as in the 
two-potential method developed by Gurvitz {\it et al.} 
\cite{Gur87,Gur88,Gur04}. 
To this end, 
we modify the core-proton potential $V(x)$ to a confining 
potential $V_{\rm mod}(x)$ 
defined as 
\begin{eqnarray}
V_{\rm mod}(x) &=& V(x) \quad |x| \leq |x_c| \nonumber \\
&=& V(x_c) \quad |x| > |x_c|. \label{eq:vmod}
\end{eqnarray}
In Fig. \ref{fig:fig1}, we show the confining potential, $V_{\rm mod}$, 
and the original potential, $V$, together with the wave function for 
the bound state of $V_{\rm mod}$. 
The position $x_c$ can be chosen arbitrarily as long as 
$V(x_c)$ is larger than the resonance energy, 
although the accuracy will be improved if $x_c$ is chosen so that $V(x_c)$ 
is as close as possible to the resonance energy\cite{Gur04}. 
When this condition is 
satisfied, the modified potential $V_{\rm mod}$ holds a bound state, which 
resembles the resonance state of the original potential. 
In this paper, we choose $x_c=10$ fm, which yields the bound state 
at $\epsilon$=2.71 MeV for the modified potential. 
The value of the modified potential at $x=x_c$ is 
$V_{\rm mod}(x_c)=4.21$ MeV, whereas the 
barrier height is $V_b = V (x=8.1~\unit{fm})$ =4.75 MeV. 

We solve the 
single-particle (s.p.) states 
of the modified \Hami, 
\begin{equation}
h_{\rm mod}(x) = -\frac{\hbar^2}{2m} \frac{d^2}{dx^2} + V_{\rm mod}(x), 
\end{equation}
as
\begin{equation}
h_{\rm mod}(x) \phi_{n}(x) = \epsilon_{n} \phi_{n}(x).
\end{equation}
In this paper, we assume that all the s.p. states with negative energy, that 
is, $\epsilon_{n} < 0$, are occupied by the core nucleus. 
Therefore, there is only one bound state in this potential at 
$\epsilon=2.71$ MeV. 
The other positive energy states are in the continuum spectra, 
which we discretize with a box of $X_{\rm box}=\pm 120$ fm.
We have confirmed that the bound state at $\epsilon$=2.71 MeV 
corresponds to 
a resonance state of the original potential, whose energy 
is stabilized against a variation of $X_{\rm box}$
\cite{HazTay70,Mai80,Zha08}. 

Using these s.p. wave functions, 
one can obtain the eigenfunctions of the modified three-body Hamiltonian, 
\begin{equation}
H_{\rm mod}(x_1,x_2) = h_{\rm mod}(x_1) + h_{\rm mod}(x_2) + v_{\rm pp} (x_1,x_2),
\label{eq:hmod}
\end{equation}
as 
\begin{equation}
\Psi_k (x_1,x_2)
= \sum_{n_1 \leq n_2} \alpha^{(k)}_{n_1n_2} \Phi_{n_1n_2} (x_1,x_2), 
\label{eq:expand}
\end{equation}
where we exclude those s.p. states occupied by the core nucleus. 
Here, $\Phi_{n_1n_2} (x_1,x_2)$ is defined as 
\begin{eqnarray}
& & \Phi_{n_1 n_2}(x_1,x_2) = \frac{1}{\sqrt{2(1+\delta_{n_1,n_2})}}
\nonumber \\
&\times & [\phi_{n_1}(x_1)\phi_{n_2}(x_2)+\phi_{n_2}(x_1)\phi_{n_1}(x_2)]
\nonumber \\
&\times & \mid S=0 \rangle . \label{eq:basis}
\end{eqnarray}
Because we use spin-independent interactions in the \Hami, 
the total spin $S$ of the two protons is a good quantum number. 
We set it to be zero (that is, the spin-singlet state). 
The spatial part of the \twop wave function is therefore symmetric 
under the exchange of $x_1$ and $x_2$. 
The coefficients $\alpha^{(k)}_{n_1n_2}$ in Eq. (\ref{eq:expand}) are 
determined by diagonalizing the 
Hamiltonian matrix for the modified Hamiltonian, $H_{\rm mod}$.

\begin{figure}[b]
\begin{center}
\includegraphics[width=\hsize]{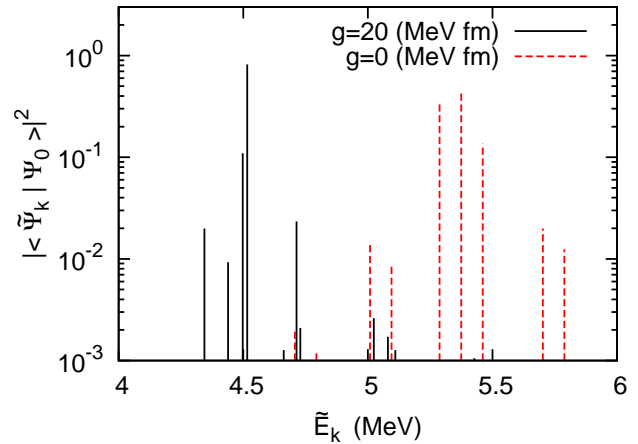}
\end{center}
\caption{(Color online) 
Overlaps between the initial state, $\Psi_0$, 
and the eigenfunctions $\tilde{\Psi}_k$ of the 
original three-body 
Hamiltonian as a function of the corresponding eigenenergies, 
$\tilde{E}_k$.
The solid and the dashed lines correspond to 
the cases of $g=20$ and $0 \; \unit{MeV} \cdot \unit{fm}$, respectively. 
Notice that the initial state is the lowest eigenstate of the 
modified Hamiltonian, $H_{\rm mod}$, with the 
eigenenergy of 4.56 MeV (5.42 MeV) for $g=20$ 
$\unit{MeV} \cdot \unit{fm}$ 
($g$=0 $\unit{MeV} \cdot \unit{fm}$). }
\label{fig:fig2}
\end{figure}

The next step is to carry out 
the time evolution for $t > 0$ starting from the lowest 
eigenfunction of the modified Hamiltonian, {\it i.e.}, 
$\Psi_{k=0}$ in Eq. (\ref{eq:expand}) at the eigenenergy $E_0$. 
That is, 
\begin{equation}
\Psi(t=0,x_1,x_2) =\Psi_0(x_1,x_2). 
\label{initst}
\end{equation}
Excluding the core-occupied states in the expansion, 
we have confirmed that there is only 
one bound state for $H_{\rm mod}$ 
in the energy region 
between 0 and $2 V_{\rm mod} (x_c)$. 
Similarly to a s.p.resonance state, 
the lowest state $\Psi_0$ corresponds to the three-body resonance state of 
the original Hamiltonian $H$. 
The energy $E_0$ corresponds to the resonance energy of the 
3-body system, that is, the $Q$-value for the \twop decay. 

We then solve 
the time-dependent Sch\"odinger equation with the original Hamiltonian $H$, 
\begin{eqnarray}
i\hbar \frac{\partial}{\partial t} 
\Psi(t,x_1,x_2) &=& H\, \Psi(t,x_1,x_2), \\
&=& (H_{\rm mod}+\Delta V) \Psi(t,x_1,x_2), 
\label{eq:tdse}
\end{eqnarray}
where 
\begin{equation}
\Delta V =V(x_1) + V(x_2) 
- V_{\rm mod}(x_1) - V_{\rm mod}(x_2), 
\end{equation}
is the difference between the original and the modified potentials. 
We expand the time-dependent \twop wave function 
with the eigenfunctions of the modified Hamiltonian, 
that is, $\Psi_k(x_1,x_2)$ given in Eq.(\ref{eq:expand}), as 
\begin{equation}
\Psi(t,x_1,x_2) = \sum_k c_k(t) \Psi_k (x_1,x_2), \label{tdexp}
\end{equation}
with the initial condition of 
\begin{equation}
c_k (t=0) = \delta_{k,0}.
\end{equation}
Substituting Eq. (\ref{tdexp}) into (\ref{eq:tdse}) and 
using the orthogonality of $\Psi_k$, 
we obtain the differential equation for 
the expansion coefficients $c_i(t)$, 
\begin{eqnarray}
i \hbar \frac{dc_i(t)}{dt} &=& \langle \Psi_i \mid H \mid \Psi \rangle \\
&=& \sum_k c_k(t) \left( E_i \delta_{i,k}
+ \langle \Psi_i \mid \Delta V \mid \Psi_k \rangle \right). 
\end{eqnarray}

Using the wave function so obtained, one 
can compute the survival probability, $P_s(t)$, and 
the decay width, $\Gamma$, as \cite{Car94,Car98,Car99A,Car99B,Car00}, 
\begin{eqnarray}
P_s(t)
&\equiv & | \langle \Psi_0 | \Psi(t) \rangle |^2 =|c_0(t)|^2 , 
\label{eq:svpb} \\
\Gamma &=& -\hbar \frac{\dot{P}_s(t)}{P_s(t)} . \label{eq:gamma}
\end{eqnarray}
When the survival probability is an exponential function of $t$, 
the decay width $\Gamma$ 
becomes a constant. 
In the next section, we will show that $P_s(t)$ indeed has an 
exponential form after a sufficient time evolution. 

\section{RESULTS}

\begin{figure}[t]
\begin{center}
\includegraphics[width=\hsize]{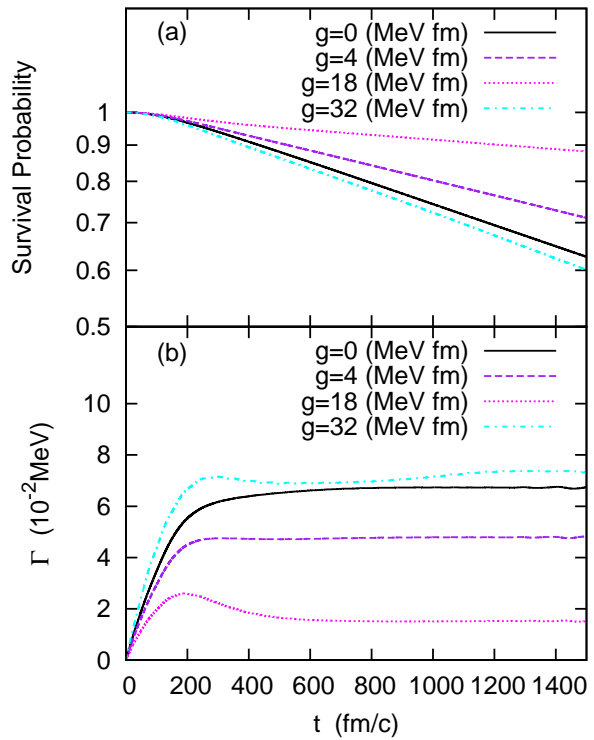}
\end{center}
\caption{
(Color online) 
(a) The survival probability as a function of time $t$ defined by
Eq.(\ref{eq:svpb}).
(b) The decay width defined by Eq.(\ref{eq:gamma}).
These are plotted for several values of $g$ indicated in the figure. }
\label{fig:fig3}
\end{figure}

\begin{figure}[t]
\begin{center}
\includegraphics[width=\hsize]{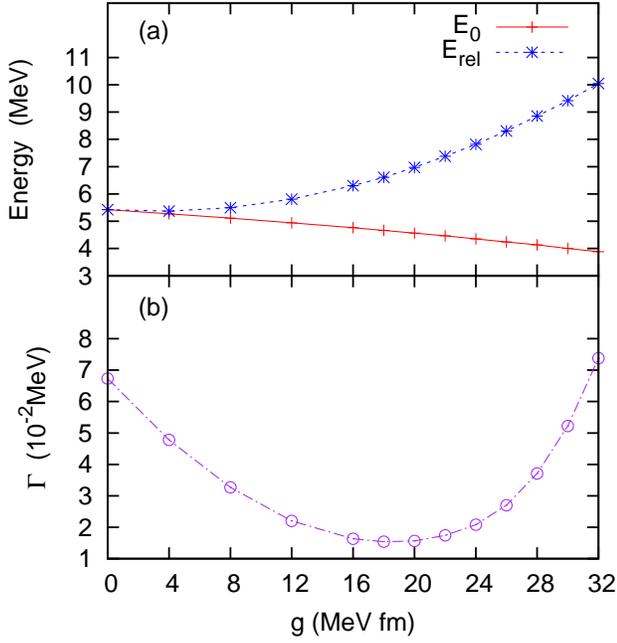}
\end{center}
\caption{
(Color online) 
(a) The decay energy $E_0$ 
 of the three-body system as a function of the 
strength of the pairing interaction, $g$. The dashed line indicates the 
asymptotic kinetic energy of a bound diproton. 
(b) The decay width estimated at $t = 1200$ fm/c. }
\label{fig:fig4}
\end{figure}

\subsection{Decay energy and width}

Before we numerically solve the time-dependent Schr\"odinger equation, 
let us first 
investigate the overlaps between the initial wave function $\Psi_0$ and the 
eigenfunctions of the original Hamiltonian $H$. 
That is, 
\begin{equation}
Q_k \equiv | \langle \tilde{\Psi}_k | \Psi_0 \rangle |^2, 
\end{equation}
where $|\tilde{\Psi}_k \rangle$ is the eigenfunctions of the 
original Hamiltonian satisfying 
\begin{equation}
H | \tilde{\Psi}_k \rangle = \tilde{E}_k | \tilde{\Psi}_k \rangle.
\end{equation}
Figure \ref{fig:fig2} shows the overlaps $Q_k$ as a function of 
$\tilde{E}_k$ for $g$=0 and 20 MeV$\cdot \unit{fm}$. 
The initial state is fragmented over several 
eigenfunctions of the original Hamiltonian, and thus forms a wave 
packet which evolves in time. 
As one can see, the fragmentation of the initial state 
is small, where the energy spreading corresponds to the decay width. 

Let us now numerically 
solve the time-dependent Schr\"odinger equation. 
To this end, we use a time mesh of $\Delta t=0.01$ fm$/c$. 
Fig.\ref{fig:fig3} shows the survival probability and the decay width 
defined as 
Eqs. (\ref{eq:svpb}) and (\ref{eq:gamma}) 
as a function of time $t$ for several values of $g$. 
One can see that the decay width converges to a constant value 
after sufficient time-evolution, 
that indicates the exponential decay-law, $P_s(t)=e^{-i\Gamma t/\hbar}$. 
Notice that the converged values for the decay width 
with this model Hamiltonian are 
in the same order as the experimental width for 
$^6$Be and $^{16}$Ne \cite{OVB89,Gri01,Gri09,Muk08}.
At shorter period, the decay width shows a transient 
behaviour \cite{Car94,Car98,Car99A,Car99B,Car00}. 
That is, 
the survival probability behaves 
like a parabolic function of $t$, whereas the decay width increases 
linearly \cite{GroKla84}. 

\begin{figure}[t]
\begin{center}
\includegraphics[width=0.9\hsize]{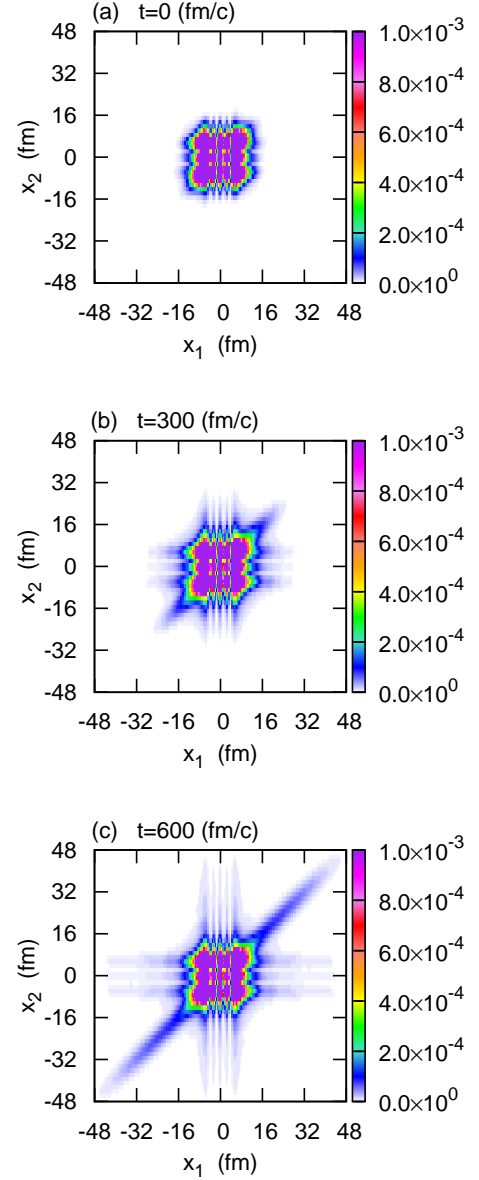}
\end{center}
\caption{
(Color online) The time-evolution of the 
density distribution $\rho (t,x_1,x_2)$ in the two-dimensional 
$(x_1,x_2)$ plane 
calculated with the pairing strength of $g$=20 $\unit{MeV} \cdot \unit{fm}$. 
The panels (a), (b), and (c) correspond to the 
density at $t=0,300$, and 600 fm/$c$, respectively. 
}
\label{fig:fig5}
\end{figure}

\begin{figure}[t]
\begin{center}
\includegraphics[width=0.9\hsize]{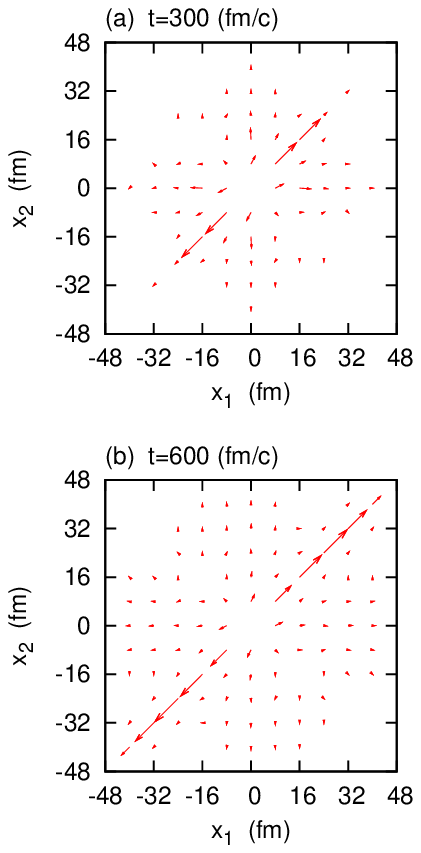}
\end{center}
\caption{
(Color online) The flux distributions 
$\bi{j} (t,x_1,x_2)=j_1(t,x_1,x_2)\bi{e}_1
+j_2(t,x_1,x_2)\bi{e}_2$, where $\bi{e}_i$ is the 
unit vector in the two-dimensional $(x_1,x_2)$ plane.  
These are obtained with $g$=20 $\unit{MeV} \cdot \unit{fm}$ 
at $t$= 300 fm/$c$ (Fig. 6(a)) and 600 fm/$c$ (Fig. 6(b)), 
and are plotted in arbitrary units.}
\label{fig:fig6}
\end{figure}

\begin{figure}[t]
\begin{center}
\includegraphics[width=0.9\hsize]{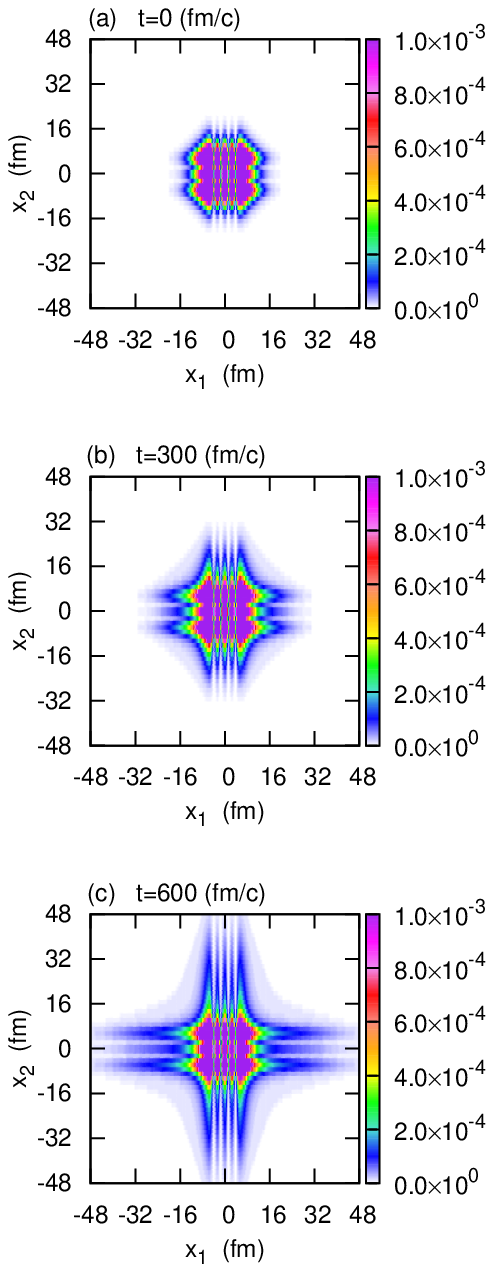}
\end{center}
\caption{
(Color online) 
Same as Fig.\ref{fig:fig5}, but for 
$g$=0 $\unit{MeV} \cdot \unit{fm}$. }
\label{fig:fig7}
\end{figure}

\begin{figure}[t]
\begin{center}
\includegraphics[width=0.9\hsize]{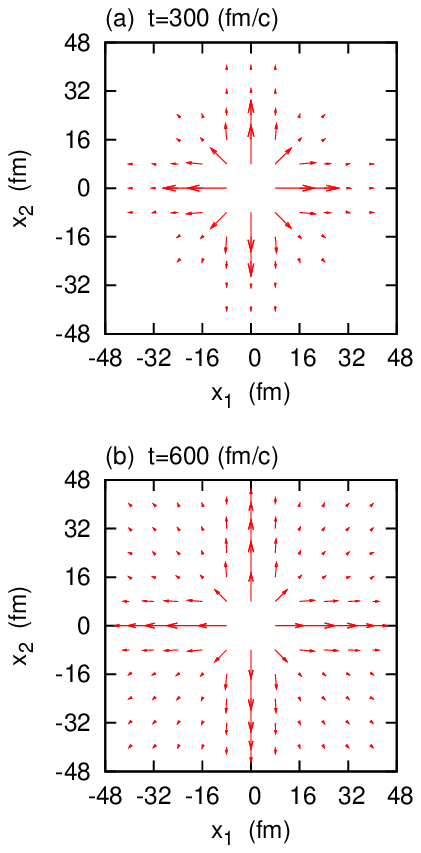}
\end{center}
\caption{
(Color online) 
Same as Fig.\ref{fig:fig6}, but for 
$g$=0 $\unit{MeV} \cdot \unit{fm}$. }
\label{fig:fig8}
\end{figure}

The dependence of the decay width on the strength of the pairing 
interaction is shown in 
Fig. \ref{fig:fig4} (b). 
We also show in 
Fig. \ref{fig:fig4} (a)
the decay energy 
$E_0$ (that is, 
the eigenenergy of the modified Hamiltonian given by Eq. \eqref{eq:hmod})
 and the asymptotic kinetic energy of diproton, $E_{\rm rel}$, defined as 
\beq
E_{\rm rel} = E_0 + B_{\rm pp}, 
\eeq
where $B_{\rm pp}=mg^2/4\hbar^2$ is the binding energy of a diproton. 
The decay width is estimated at 
$t = 1200$ fm/$c$, where it has been well converged (see Fig. 
\ref{fig:fig3}). 

The decay width $\Gamma$ first decreases as a function of $g$, 
despite that the diproton kinetic energy $E_{\rm rel}$ increases. 
This should be related to the decrease of the decay energy $E_0$, 
indicating that the sequential two-proton emissions is the 
main decay mechanism in this region of $g$ 
even though two protons are bound in 
this one-dimensional model. 
For $g\geq 18$ MeV$\cdot \unit{fm}$, 
on the other hand, the decay width increases. 
This is consistent with the increase of $E_{\rm rel}$, 
suggesting that 
the direct diproton decay, that is the emission of a deeply bound 
diproton, is the main mechanism in this region. 

The transition from the sequential to the diproton decays 
 will be clarified more in the next subsection.

\subsection{Two-particle density and flux distributions}

In order to confirm a transition from a sequential to a simultaneous decays 
discussed in the previous subsection, 
we next discuss the time evolution of two-particle density 
distribution, 
\begin{equation}
\rho (t,x_1,x_2) = \mid \Psi(t,x_1,x_2) \mid ^2.
\end{equation}
We also analyze the flux distribution defined as, 
\begin{equation}
j_i(t,x_1,x_2)= \frac{\hbar }{2im}
\left( \Psi^* \frac{\partial \Psi}{\partial x_i}
- \frac{\partial \Psi^*}{\partial x_i} \Psi \right)~~~~~(j=1,2).
\end{equation}
Note that the \twop density is normalized as 
\begin{equation}
\int_{-\infty}^{\infty}\rho (t,x_1,x_2) dx_1 dx_2 = 1.
\end{equation}
Figs.\ref{fig:fig5} and \ref{fig:fig6} show 
the two-particle density 
and flux distributions, respectively, 
for 
$g=20 \; \unit{MeV} \cdot \unit{fm}$ 
at 
$t=0, 300$ and $600$ fm/$c$. 
The corresponding quantities for 
$g=0$ are also shown in 
Figs.\ref{fig:fig7} and \ref{fig:fig8}. 
The flux distributions are plotted in a form of vector 
at each value of ($x_1,x_2$) 
in the two-dimensional ($x_1,x_2$) plane, where 
the core nucleus is located at the origin. 
Note that 
there is no flux distribution at $t=0$ 
because the initial wave function $\Psi (t=0)$ 
can be taken to be real, and we do not show it in the figures. 

Figs. \ref{fig:fig5}(a) and \ref{fig:fig7}(a) show 
the density distribution for the initial \twop state, 
which is confined within the modified 
potential, $V_{\rm mod}$. 
Because of the pairing correlation, 
the initial density for $g=20 \; \unit{MeV} \cdot \unit{fm}$ has an 
asymmetric form, 
with the peaks along $x_1 = x_2$ being higher than 
those along $x_1 = -x_2$\cite{HVPS11}. 
The peaks along the 
$x_1 = x_2$ line, that is, in the first and third 
quadrants of these panels, 
correspond to 
a compact diproton cluster. 
On the other hand, 
the peaks along the $x_1 = -x_2$ line ({\it i.e.,} in the second 
and the fourth quadrants) 
correspond to a configuration in which two protons are located 
opposite to the core nucleus. 
If we discard the pairing interaction, 
the density distribution has 
four symmetric peaks, as shown in Fig. \ref{fig:fig7}, 
that is, the probability in the 
first and third quadrants 
is the same as that in 
the second and fourth quadrants 
\cite{HVPS11}. 

The effect of pairing correlation is apparent also during the 
time evolution. 
In the presence of the pairing correlation, the extension of 
the two-particle density along the $x_1=x_2$ line increases significantly, 
although the extension along the $x_1=0$ or $x_2=0$ lines 
is not negligible. 
This is in marked contrast with the uncorrelated case shown in Fig. 
\ref{fig:fig7}, in which the two-particle density expands democratically. 
That is, in the uncorrelated case, the probability of emission of the two 
protons in opposite directions is equal to that in the same direction. 
The flux distribution shown in Figs. \ref{fig:fig6} and \ref{fig:fig8} 
also indicate the same behaviour. 

\begin{figure}[t]
\begin{center}
\includegraphics[width=\hsize]{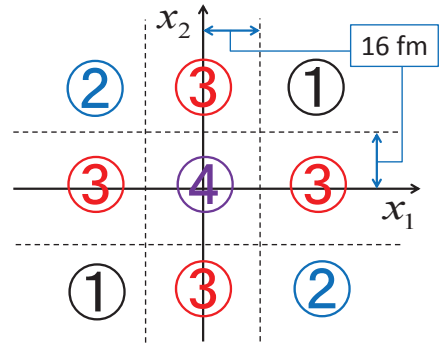}
\end{center}
\caption{(Color online) 
The four regions in the $(x_1,x_2)$ plane 
used to calculate the partial probabilities shown in 
Fig. \ref{fig:fig10}. 
The boundaries of each region are at $x_1=\pm$16 fm and 
$x_2=\pm$16 fm. } 
\label{fig:fig9}
\end{figure}

\begin{figure}[t]
\begin{center}
\includegraphics[width=\hsize]{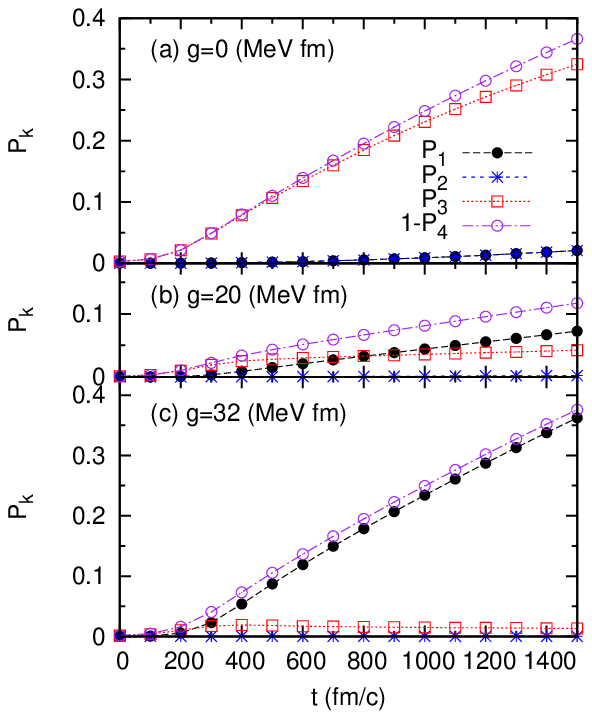}
\end{center}
\caption{(Color online) 
The time-evolution of the partial probabilities 
in the regions defined in Fig. \ref{fig:fig9}
for $g$=0, 20, and 32 MeV$\cdot$fm. 
The total decay probability, 
$P_{\rm tot}\equiv1-P_4=P_1+P_2+P_3$, is also shown 
by the dot-dashed lines. }
\label{fig:fig10}
\end{figure}

In order to investigate the time evolution more quantitatively, 
we divide the $(x_1,x_2)$ plane into four regions shown in 
Fig. \ref{fig:fig9}. 
That is, 
(i) the region of $x_1 > $ 16 fm and $x_2 > $ 16 fm, as well as 
the region of $x_1 < -16$ fm and $x_2 < -16$ fm, 
(ii) the region of $x_1 > $ 16 fm and $x_2 < -$ 16 fm, as well as 
the region of $x_1 < -16$ fm and $x_2 > 16$ fm, 
(iii) 
the region of $-16 \leq x_1 \leq $ 16 fm and $|x_2| > 16$ fm, 
as well as 
the region of $-16 \leq x_2 \leq $ 16 fm and $|x_1| > 16$ fm, 
and (iv) the rest in the $(x_1,x_2)$ plane. 
At each time, 
we integrate the two-particle density distribution in each region, 
\begin{equation}
P_k (t) = \int_{\rm region~k} \rho (t,x_1,x_2) dx_1 dx_2 , \; (k = 1 \sim 4).
\end{equation}
The time evolution of 
these partial probabilities 
is shown in 
Fig.\ref{fig:fig10} for 
$g = 0, 20$ and 
$32 \; \unit{MeV} \cdot \unit{fm}$. 
We also show the total decay probability, that is, 
$1-P_4=P_1+P_2+P_3$. 

We first discuss the behaviour for the uncorrelated case shown in 
Fig. \ref{fig:fig10} (a). 
In this case, the dominant process  
is the decay into the third region, $P_3$, which 
corresponds to an emission of one of the valence protons while 
the other proton remains 
inside the core-proton potential. 
As there is no active 
s.p. bound state in the present core-proton potential, 
the second proton mainly occupies the s.p. resonance state. 
This resonance state eventually decays and the second proton 
is emitted outside the potential after a sufficient time-evolution. 
This is nothing but the sequential two-proton decay, and 
can be clearly seen in the probabilities $P_1$ and $P_2$, 
that exist only at $t\gtrsim 600$ fm/$c$. 
Notice that $P_1$ and $P_2$ are identical to each other, 
since the second proton is emitted 
into either 
the left or the right direction with respect to the core nucleus 
with an equal probability irrespective to the position of the 
first proton. 

For $g=20 \; \unit{MeV} \cdot \unit{fm}$ shown in 
Fig. \ref{fig:fig10}(b), 
the probability in the region (i) increases considerably due to 
the pairing correlation, while $P_3$ decreases significantly. 
This partly corresponds to an emission of a bound diproton, that is, 
the simultaneous two-proton decay. 
Notice, however, that $P_3$ is still larger than $P_1$ 
at $t\lesssim$ 800 fm/$c$, and a sequential decay also coexists for this 
value of $g$. 
As we have shown in Fig.\ref{fig:fig4}, the total decay probability, 
$P_1+P_2+P_3$, decreases compared to the uncorrelated case. 

When the pairing interaction is even stronger, 
the simultaneous diproton decay becomes dominant. 
See Fig.\ref{fig:fig10} (c) 
for $g=32 \; \unit{MeV} \cdot \unit{fm}$. 
In this case, $P_1$ is the dominant part of the total 
decay probability, except for the short time region, at which 
the high energy components in the initial wave function 
quickly escape from the potential barrier. 
The long-time behaviour in this case 
may correspond to alpha decays in realistic nuclei, 
for which a tightly bound alpha particle tunnels through 
the Coulomb barrier of the daughter nucleus. 

From these studies, it is evident that the present 
one-dimensional three-body model nicely describes a transition from 
an uncorrelated case to a strongly correlated case for many-particle 
tunneling decays. 

\section{SUMMARY}

We have employed the 
time-dependent method and investigated 
many-particle tunneling decays, particularly 
the two-proton radioactivity. 
To this end, 
we have used 
a one-dimensional three-body model which 
consists of a core nucleus and two valence 
protons. 
In order to describe the decaying process, 
we first confined the two-proton wave function inside 
a confining potential. The confining potential was then 
changed to the original potential, with which the two-proton 
wave function evolves in time. 
We have confirmed that the survival probability 
follows 
the exponential decay-law after a sufficient time-evolution, 
yielding a constant decay width. 
We have found that 
an emission of a bound diproton is enhanced due to the 
pairing correlation, as was evidenced in 
the time evolution of the 
density and flux distributions. 
We have also analyzed the partial probabilities 
and discussed the relative importance of the sequential and 
simultaneous two-proton decays. 
We have shown that, for the uncorrelated case, the sequential 
decay is the dominant decay process, while the simultaneous decay 
plays an essential role in the case of strong pairing 
correlation. For an intermediate value of the pairing 
strength, we have shown that both the simultaneous and the 
sequential two-proton emission coexist. 

The one-dimensional three-body model which we employed in 
this paper is a simple schematic model, with which a deep understanding 
of many-particle decay process may be obtained. 
One drawback, however, is that two protons are inevitably bound 
even with an infinitesimal attraction between the two protons. 
Even though our model nicely demonstrates the 
coexistence of the simultaneous and sequential decays for an 
intermediate pairing interaction, in reality two protons are never 
bound in vacuum. 
It will be an important task to extend the present study to 
realistic two-proton emitters in three-dimension, such as 
$^6$Be and $^{16}$Ne nuclei. 
A work towards this direction is in progress, and we will 
report on it 
in a separate publication. 

\begin{acknowledgments}
We thank D. Lacroix for useful discussions. 
This work was supported by the Global COE Program
``Weaving Science Web beyond Particle-Matter Hierarchy'' at
Tohoku University,
and by the Japanese
Ministry of Education, Culture, Sports, Science and Technology
by Grant-in-Aid for Scientific Research under
the program number (C) 22540262.
\end{acknowledgments}

\end{document}